\documentclass{elsart}

\usepackage{natbib}

\usepackage{graphicx}

\usepackage{amssymb}
\usepackage{latexsym}

\newcommand{\bfR}{\mathbf{R}}
\newcommand{\bfv}{\mathbf{v}}
\begin{document}

\begin{frontmatter}



\title{Are vertical cosmic rays the most suitable to radio detection ?}


\author{Thierry Gousset}, 
\author{Olivier Ravel}, 
\author{Christelle Roy\corauthref{cor}}
\corauth[cor]{Corresponding author.}
\ead{Christelle.Roy@subatech.in2p3.fr} 

\address{SUBATECH, Laboratoire de Physique Subatomique et des Technologies
Associ\'ees\\
UMR Universit\'e de Nantes, IN2P3/CNRS, Ecole des Mines de Nantes\\
4, rue Alfred Kastler, 44307 Nantes cedex 3, France}

\begin{abstract}
The electric field induced by extensive air showers generated by high
energy cosmic rays is considered and, more specifically, its
dependence on the shower incident angle. It is shown that for
distances between the shower axis and the observation point larger
than a few hundred meters, non-vertical showers produce larger fields
than vertical ones. This may open up new prospects since, to some
extent, the consideration of non-vertical showers modifies the scope of
the radio-detection domain.
\end{abstract}


\end{frontmatter}

\section{Introduction}
\label{Introduction}
The observation of ultra high-energy cosmic rays ($E\geq 10^{18}$~eV)
is nowadays a question of paramount importance. They can be detected
indirectly via the observation of the Extensive Air Showers (EAS) they
create while interacting with the atmosphere. The induced secondary
particles generate a radio-electric field whose measurements may
provide an alternative to the experimental techniques consisting of 
counting the shower particles collected in ground detectors. It turns
out that the radio signal becomes significant provided that an
efficient charge separation mechanism occurs~\cite{allan}.

The radio frequency component associated with EAS was studied in the
1960's. At that time, the purpose was mainly to demonstrate the observability of such a
radio-electric field~\cite{askaryan,jelley}. The strategy was then to
detect a radio pulse in coincidence with particles in an experimental
set-up consisting of one antenna located as close as possible to some
particle detectors. Such a device selects mainly \emph{vertical}
showers.

The purpose of this letter is to stress that, in the case of not-too-close
impact parameters (i.e., large distances between the shower axis and
the observation point), vertical showers produce a radio-electric
signal weaker in amplitude than oblique ones. The reason is the
ultra-relativistic character of most of the shower particles that
leads to a strong forward enhancement of the electric field. For a
given impact parameter, this means that the point at which a charge
produces the largest signal along its trajectory is located far above
the point of closest approach. This forward-peaked structure is
counterbalanced by the rise in the number of charges as the shower
develops in the dense layers of atmosphere. Quantitatively, as
explained in Sect.~\ref{sec:calc}, the combination of both effects
will favor the observation of \emph{non-vertical} showers. 

Sect.~\ref{sec:outlook} unfolds the consequence of this
observation. The point is that, using a large array of scattered
antennas in order to compensate for the scarcity of high-energy cosmic
rays, there is some interest in examining the possibility of radio
detection of EAS at a somewhat larger impact parameter. According to the
above statement, such an apparatus, therefore, puts emphasis on the
detection of non-vertical hadron-induced showers. 

\section{Effect of shower incidence on the radio signal}
\label{sec:calc}

This section aims at studying the dependence of the radio frequency
signal on the incidence angle of the air shower. For this purpose, one
of the mechanisms possible for electromagnetic field generation has
been chosen, namely the synchrotron radiation. The latter is
associated with the deviation, due to the geomagnetic field, of the EAS
charged particles. The specific choice of synchrotron radiation should
not be considered as a limitation, since the main lines of
argumentation are generic and can be transposed to any other
mechanism. The additional choices concern the nature and the incident
energy of the cosmic rays : a $10^{18}$~eV proton induced-shower will
be considered throughout.

The first element that drives the electromagnetic field emitted during
the development of the air shower is obviously the evolution of the
number of charges as a function of time. The way it affects the
expression of the electric field depends on the details of the
mechanism considered. However, the main feature is the fact that, this
charge number rises from a few units to several $10^8$, and that, this
growth takes place on a characteristic depth scale of one
atmosphere~\cite{gaisser}. The consequence for the time evolution of the signal is
analyzed in Sect.~3 of Ref.~\cite{allan}. At zero impact parameter, the
above time sequence is Doppler contracted to almost zero and the
signal duration results from the longitudinal lag and lateral spread
of the shower core. On the contrary, for impact parameters larger than
a few hundred meters, the Doppler contraction being less important, it
is the shower growth and decay which govern the time behavior of the
radio signal. In this configuration, the extension of the shower core
only introduces weak time dispersions to the signal. As a consequence,
it is reasonable for an estimate of the electric field at large impact
parameter to modelize the shower as a point-like system moving along a
straight line and containing a number of electrons and positrons 
varying with time. 

The second element which influences the total electric field is the
space-time dependence of the field of a single charge. It should be
noted that the interest is not so in following a specific charge which
suffers deviation in the earth magnetic field, and in addition loses
energy before disappearing from the shower core. Rather, since the
total field is the sum over the set of charges, the interest is in the
field generated by a representative of particles of a given energy 
sitting where most particles sit, i.e., in the shower core. In line
with the above model for the shower core, such a representative has its
speed directed along the shower axis at any time.

Before combining both elements, the single-charge electric field is
calculated in a first step. The electric field created by an
accelerated charge $e$ at a given observation point $A$ at time $t$
reads
\begin{equation}
\mathbf{E}(t,A)=\frac{e}{4\pi\epsilon_0 c_0}\left[
\frac{\bfR\wedge\Bigl[(\bfR-nR\bfv)\wedge\dot\bfv\Bigr]}%
{|R-n\,\bfv\cdot\bfR|^3}\right]_{t'},
\end{equation}
where $\bfv$ is the particle speed in units of $c_0$, the light
velocity in vacuum; $\dot\bfv$ is the time derivative of $\bfv$; $n$
is the refractive index of the medium and $\bfR=\overrightarrow{QA}$
where $Q$ is the location, at time $t'$, of the particle. $t$, $t'$
and $A$ are linked by\footnote{In the
\v Cerenkov regime, this equation may have two solutions for $t'$. In
this case, both have to be taken into account for calculating the
field defined by Eq.~(1).}
\[
\frac{c_0}{n}(t-t')=R=||\overrightarrow{QA}||.
\]

A first remark about Eq.~(1) is that two opposite charges $+e$ and
$-e$ have opposite accelerations $+\dot\bfv$ and $-\dot\bfv$ so that
their respective electric fields add up. This is the manifestation for
the case of synchrotron radiation of the charge separation mechanism
mentioned in the introduction. 

The form of the field in Eq.~(1) assumes a signal propagation in a
medium of constant $n$. It is shown in Ref.~\cite{allan} that the time
variation introduced in the signal when considering refractive index
effects is small at large impact parameters, i.e., well outside of the
\v{C}erenkov cone.

\begin{figure}
\begin{center}
\includegraphics[width=7cm]{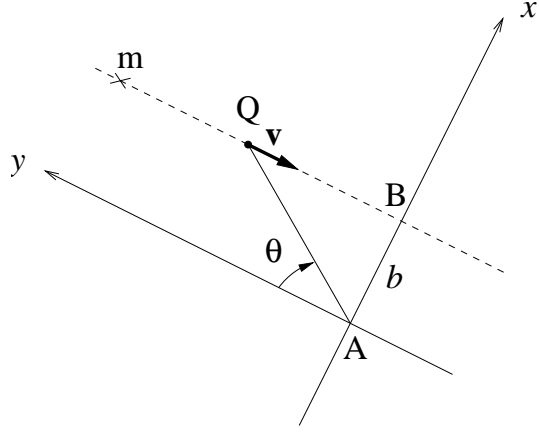}
\caption{
Geometrical quantities used in Eq.~(2). The dashed line is the
shower axis. At time $t^\prime$, the particle $Q$ may be located with
$\theta$, the angle between $\protect\overrightarrow{Ay}$ and
$\protect\overrightarrow{AQ}$. $B$ is the point of closest approach, whereas $m$ is 
the point where the quantity in Eq.~(2) is maximal.\label{fig1}}
\end{center}
\vspace*{0.4cm}
\end{figure}

In order to understand the variation of $\mathbf{E}$ in Eq.~(1), a 
special attention has to be paid to the denominator, cubic power of
the term $|R-n\,\bfv\cdot\bfR|^{-1}$. This factor combines
the $R^{-1}$ term, related to the distance between the charge and the
observation point, with the forward-angle-peaked term 
$|1-n\,\bfv\cdot\bfR/R|^{-1}$ whose effect becomes more
pronounced when the particle velocity approaches $c=c_0/n$. The
dependence on the angle can be shown explicitly, using
the geometrical quantities defined on Fig.~\ref{fig1}, and reads :

\begin{equation}
\frac{1}{|R-n\,\bfv\cdot\bfR|}
=\frac{1}{b}\frac{\sin\theta}{|1-nv\cos\theta|}.
\end{equation}

\begin{figure}
\begin{center}
\includegraphics*[width=8cm]{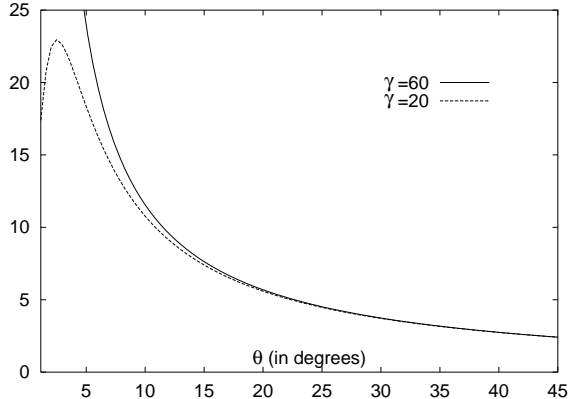}
\caption{$\sin\theta/(1-nv\cos\theta)$ as a function of
  $\theta$. Two cases are represented : $v<1/n$ (Lorentz factor $\gamma=20$)
  and $v>1/n$ ($\gamma=60$).\label{fig2}}
\end{center}
\vspace*{0.4cm}
\end{figure}

The evolution of this term as a function of $\theta$ is shown on
Fig.~\ref{fig2}.  For a given impact parameter, this factor is a
decreasing function of $\theta$ in the range $\theta>\theta_m$. For
particle velocities $v<1/n$ (in units of $c_0$), $\theta_m$ is defined as : 

\[
\theta_m=\arccos(nv).
\]

For $v>1/n$, it is given by : 
\[
\theta_m=\arccos\left(\frac{1}{nv}\right), 
\]

and corresponds to the definition of the \v Cerenkov
angle.\footnote{For $v>c$, the factor in Eq.~(2) diverges at
$\theta=\theta_m$ and the analysis should be reformulated to account
for the non constancy of the refraction index. This complication is
however unnecessary when limiting the discussion to large impact
parameters, since $\theta=\theta_m$ corresponds to points at high
enough altitudes where the cosmic ray has not yet showered.} In both
cases, $\theta_m$ is a very small angle as long as ultra-relativistic
particles are concerned. The consequence is that the factor
$|R-n\,\bfv\cdot\bfR|^{-1}$ reaches its maximum value at the point
$m$, such that $\theta(m)=\theta_m$, far above the point of closest
approach $B$ (see Fig.~\ref{fig1}). As an example, the distance
between $m$ and $B$ amounts to $23\times b$ for a particle with
$\gamma=20$ and $56\times b$ for a particle with $\gamma=60$.

Thus, if the $\theta$ dependence in Eq.~(1) was contained only in the
denominator, the point of maximal brightness for the electric field
would be reached when the charge is at $m$ (not at $B$).  As a matter
of fact, the numerator in Eq.~(1) is also a function of
$\theta$. Moreover, it depends on the orientation of the particle
acceleration. In the case where $v>c$, this does not modify the fact
that the electric field is a decreasing function of $\theta$ in the
range $\theta>\theta_m$. Hence, for clarity, discussion will be
restricted to this particular case. For particle velocities $v<c$, the
discussion requires a case study that will not be examined in this
letter. It can nevertheless be verified that the general conclusion
derived below also applies to this second case.

\begin{figure}
\begin{center}
\includegraphics[width=8cm]{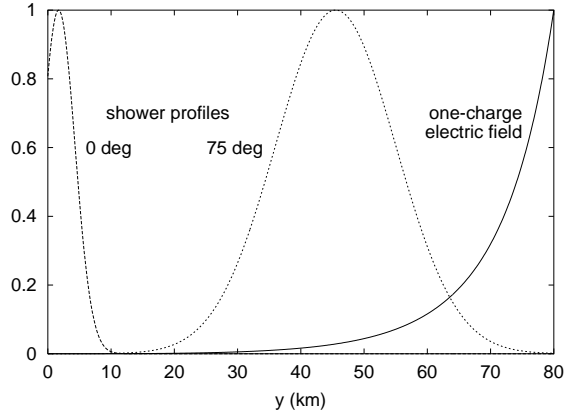}
\caption{Charge profiles along the shower axis $y$ for both vertical
  (0$^\circ$) and tilted (75$^\circ$) proton-induced showers. The electric
  field shape of a single charge is also represented as a function of
  $y$. Calculations have been performed for $b=2$~km and
  $\gamma=60$.
\label{fig3}}
\end{center}
\end{figure}

After having considered a unique charge, calculations are carried out
for the full set of charges, in order to take into account the shower
development in the atmosphere. The field in Eq.~(1) is thus multiplied
by a number of charges following a given time profile. Fig.~\ref{fig3}
displays the various quantities which have to be combined, as a
function of the position along the shower axis $y = - v \times t'$:
The electric field induced by a single charge and two different charge
profiles corresponding to shower incident angles fixed at $\alpha =
0^\circ$ and $\alpha = 75^\circ$. Though this is not shown on
Fig.~\ref{fig3}, it is easy to guess that the multiplication of the
0$^\circ$ profile with the single-charge electric field peaks at a
much smaller amplitude than the multiplication of the 75$^\circ$
profile with the same field. To be specific, the ratio of maximal
values is in this case $1:4000$. Such a ratio depends on the
parameters chosen, and on the simplifications adopted in the present
derivation. However, the conclusion remains that when the
single-charge electric field has its maximum at large $y$, i.e., for
large impact parameters, inclined showers produce larger electric
field than a vertical one.

\begin{figure}
\begin{center}
\includegraphics[width=7cm]{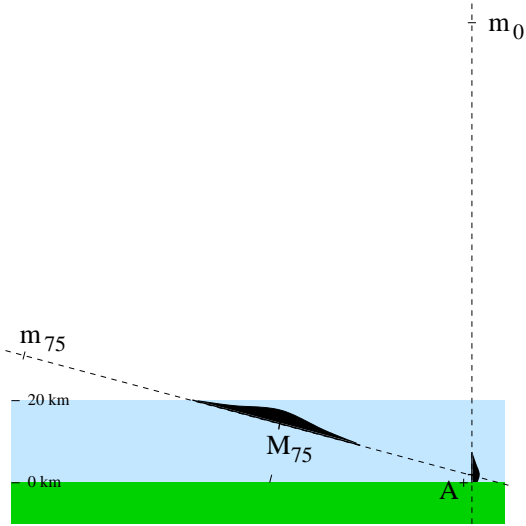}
\caption{Illustration of air shower developments in the atmosphere, 
for two different shower incidence angles ($\alpha = 0^\circ$ and
$\alpha=75^\circ$ defined with respect to the vertical direction).\label{fig4}}
\end{center}
\vspace*{0.4cm}
\end{figure}

Another way of picturing the above discussion is shown on
Fig.~\ref{fig4}. The development of air showers, with a maximum at
$M_\alpha$, for either $\alpha=0^\circ$ or $\alpha=75^\circ$ incident
angle is illustrated. Also shown in both cases is the location of the
maximal single-charge field ($m_0$ and $m_{75}$) for impact parameter
$b=2$~km. The tilted shower is more efficient for radiation than the
vertical one because $M_\alpha$ is closer to
$m_\alpha$ in the former case. For large impact parameters, this
always occurs in a non-vertical configuration. Let us be more precise
about what ``large'' means. Fig.~\ref{fig4} suggests that, for a
shower of incidence $\alpha$, the best configuration is obtained when
$M_\alpha=m_\alpha$. Furthermore, $M_\alpha$ moves away from the point
of closest approach, $B$, as $\alpha$ increases while $m_\alpha$ moves
away from $B$ as the impact parameter, $b$, increases. Choosing now an
intermediate angle value of $45^\circ$, large impact parameters
correspond to those greater than $b_0$ defined by 
$m_{45}(b_0)=M_{45}$. Typically, for a $10^{18}$~eV shower, $M_{45}$
is located at 5~km ahead of $B$. This leads to $b_0=0.1$~km for
$\gamma=60$ and $b_0=0.2$~km for $\gamma=20$.

\section{Outlook}
\label{sec:outlook}

The above results put emphasis on horizontal air
showers. The latter, usually considered in the study of energetic 
neutrino cosmic rays, appear to be also relevant for 
the domain of radio-detection. Thus, the perspectives of
radio-detection are somewhat enlarged if (quasi-)vertical showers are
not the only ones of interest.

For ground detection of secondary particles, air showers at large
incident angle are the hardest to detect if only because their main
components cannot reach the ground. The situation is completely
different in the case of radio detection. In order to take full
advantage of this fact, it is necessary to give up standard trigger
methods that use particle ground detection. One should rather use the
radio frequency signal itself as a trigger. Such a possibility is at
the moment tested by the CODALEMA experiment~\cite{codalema}: Triggering 
is performed on transients whose levels exceed a given threshold, and
in a frequency band-width where strong steady radio transmitters are
absent. Alternative attempts using coincidence between different
electric pulses collected on a few antennas have also been
experimented~\cite{castagnoli}.

According to Sect.~\ref{sec:calc}, such a system is more efficient for
radio-detection of non-vertical showers as soon as impact parameters
larger than a few hundred meters become detectable. With this proviso
in mind, this means that, for a given cosmic ray incident energy, the
accessible region of impact parameters is larger than expected.
Conversely, the detectable cosmic ray energy is lower, and a larger
rate of triggers is expected, for non-vertical showers with a given
impact parameter in this range. Calculations of Sect.~\ref{sec:calc}
are, however, too limited to estimate the gain in sensitivity for the
detection of such events. 

Finally, some comments on antenna networks are in order. A key
question in the observation of very low cosmic ray flux, is the
surface achievable for a given number of antennas, and implicitly,
that of antenna spacing. The present study has some relevance in terms
of antenna network capabilities regarding horizontal versus vertical
air showers, since the non-vertical configuration allows for a wider
spacing, or vice versa, for lower incident energies and a given
spacing.

The question of antenna network has been considered by the CASA/MIA
collaboration~\cite{casa}: There could be a possibility of equipping
AUGER~\cite{auger} with antennas, thereby allowing for a detection in
coincidence between particles collected in ground detectors and
registered radio pulses. The above considerations may open up somewhat
new perspectives to such a project.

\section*{Acknowledgments}
We thank R. Dallier, F. Haddad, P. Lautridou and K. Werner for
helpful and motivating discussions.





\end{document}